\documentclass[runningheads]{llncs}
\usepackage{graphicx}
\usepackage{amsmath}

\usepackage{bera}
\usepackage{listings}
\usepackage{xcolor}

\usepackage{tikz}

\colorlet{punct}{red!60!black}
\definecolor{forestgreen}{HTML}{229954}
\definecolor{background}{HTML}{EEEEEE}
\definecolor{delim}{RGB}{20,105,176}
\colorlet{numb}{magenta!60!black}

\lstdefinelanguage{json}{
    basicstyle=\normalfont\ttfamily,
    numbers=left,
    numberstyle=\scriptsize,
    stepnumber=1,
    numbersep=8pt,
    showstringspaces=false,
    breaklines=true,
    frame=lines,
    escapechar={|},
    backgroundcolor=\color{background},
    literate=
     *{0}{{{\color{numb}0}}}{1}
      {1}{{{\color{numb}1}}}{1}
      {2}{{{\color{numb}2}}}{1}
      {3}{{{\color{numb}3}}}{1}
      {4}{{{\color{numb}4}}}{1}
      {5}{{{\color{numb}5}}}{1}
      {6}{{{\color{numb}6}}}{1}
      {7}{{{\color{numb}7}}}{1}
      {8}{{{\color{numb}8}}}{1}
      {9}{{{\color{numb}9}}}{1}
      {:}{{{\color{punct}{:}}}}{1}
      {,}{{{\color{punct}{,}}}}{1}
      {\{}{{{\color{delim}{\{}}}}{1}
      {\}}{{{\color{delim}{\}}}}}{1}
      {[}{{{\color{delim}{[}}}}{1}
      {]}{{{\color{delim}{]}}}}{1},
}

\begin{document}

\title{Open Government Data Corpus for Table Search}
%
%
\author{Michael Glass\orcidID{0009-0000-1505-4667} \and
Sugato Bagchi\orcidID{0009-0005-1173-600X} \and \\
Oktie Hassanzadeh\orcidID{0000-0001-5307-9857} \and \\
Gaetano Rossiello\orcidID{0000-0003-1042-4782} \and
Alfio Gliozzo\orcidID{0000-0002-8044-2911}}
\authorrunning{M. Glass et al.}
%
\institute{IBM Research AI, Thomas J. Watson Research Center \\ 
    Yorktown Heights, NY \\
\email{\{mrglass,bagchi,hassanzadeh,gliozzo\}@us.ibm.com}\\
\email{gaetano.rossiello@ibm.com}}
\maketitle              
\begin{abstract}
Increasing amounts of structured data can provide value for research and business if the relevant data can be located. 
Often the data is in a data lake without a consistent schema, making locating useful data challenging.
Table search is a growing research area, but existing benchmarks have been limited to \textit{displayed} tables. Tables sized and formatted for display in a Wikipedia page or ArXiv paper are considerably different from data tables in both scale and style.  By using metadata associated with open data from government portals, we create the first dataset to benchmark search over data tables at scale.  We demonstrate three styles of table-to-table related table search.  The three notions of table relatedness are: tables produced by the same organization, tables distributed as part of the same dataset, and tables with a high degree of overlap in the annotated tags.  The keyword tags provided with the metadata also permit the automatic creation of a keyword search over tables benchmark.  We provide baselines on this dataset using existing methods including traditional and neural approaches. 
\end{abstract}
\section{Introduction}

For many scientific and business purposes, the size of the available data is beyond the point where it can be manually cataloged or even managed with a consistent metadata schema.  Rather than attempting to impose structure a priori, and systematize data into a database, recent approaches build on a data lake.  A data lake permits storing diverse structured and unstructured data and uses a scalable infrastructure to ensure efficient access.  Effective access requires tools for data discovery.

An initial stage of data discovery begins with a user intent. This intent could be presented as a natural language description, but a more common interface is to provide keywords for search.  The user may provide many keyword queries, gathering a set of data tables relevant to their intent.  This set of tables can then be expanded through table-to-table search.  This includes search for joinable and unionable tables, but also less precise notions of table relatedness such as topic similarity.



\section{Related Work}
\label{sec:related_work}

Previous efforts at table search have considered tables collected from pages meant for display.  Web Table Retrieval \cite{chen2021wtr} uses HTML tables from Common Crawl and uses crowd sourcing and query logs to gather 60 queries.  These queries are then attempted over several baseline methods and the resulting tables are pooled for relevance judgements.  Crowd sourcing is then used to provide over 6000 query-table relevance judgements.

Web Table Retrieval follows previous work on a Wikipedia table search benchmark~\cite{zhang2018ad}.  The same 60 queries were used and again a pooled evaluation provides relevance judgements for over 3000 query-table pairs.  In this work a majority opinion of three trained judges is the relevance judgement.
A common issue in pooled evaluation benchmarks is that the set of candidates is drawn from existing search methods. Therefore methods that improve recall without improving the rank ordering of tables retrieved by other methods will not be well assessed.

TableArXiv \cite{gao2017scientific} introduced a benchmark for keyword search over tables with the tables drawn from scientific Physics-related pre-prints on arXiv.  Students majoring in Physics formulated 105 queries and provided relevance judgements for up to 100 tables per query.

Keyword search benchmarks can be used as related table benchmarks by the assumption that if two tables are relevant for the same query they are related.  TabSim~\cite{tabsim} applies this approach to TableArXiv and introduces a related table benchmark dataset based on tables in PubMed Central (PMC) articles.  A sample of 150 tables are used as queries and 10 result tables for each query are given relevance judgements.

Another approach to a related table benchmark dataset is TableNet~\cite{tablenet}.  TableNet considers two relations \texttt{subPartOf} and \texttt{equivalent}.  The tables are extracted from Wikipedia and crowd sourcing is used to find relations between candidate table pairs obtained by heuristic filtering.  

The existing table search benchmarks have key limitations: 1) tables extracted from pages necessarily have a relatively small number of columns and rows, 2) the small number of queries limits statistical power, and 3) the benchmarks are constructed as pooled evaluations, which limits the assessment of methods that can increase recall.  Due to the limitations of existing table search benchmarks, research exploring search in data lakes typically resorts to user studies \cite{aurum}, measures of runtime performance \cite{josie}, or impact in downstream tasks \cite{juneau}.

Sarma et al. \cite{findingRelatedTables} defines related tables as a pair of tables that can form a single \textit{coherent} table. And therefore the two original tables are recoverable from the combined table by a sequence of selection and projection operations. Note that this is different from our notion of relatedness. 


\section{CKAN}

Open government data is data released by various governments to facilitate research or just provide transparency.  It is accessible through CKAN (Comprehensive Knowledge Archive Network) which provides an API to download the data.  The data in CKAN is organized in three tiers: site, dataset and table. First a site or catalog collects data from a particular source, in our case a government. Then the data is organized into datasets which collect related data tables.  

We downloaded data from seven English language open government portals.  Table \ref{tab:ckan_stats} gives statistics about the size of each catalog.
\begin{table*}[!htb]
\caption{CKAN dataset statistics} 
\label{tab:ckan_stats}
\centering
\begin{tabular}{l|rrrrrr}
\hline
Catalog & Orig. GB &	Datasets & Resources &  Dataframes & Avg. Cols& Avg. Rows \\
\hline
catalog.data.gov & 243 & 36390 & 67966 & 46886 & 24.6 & 20264.9 \\
open.canada.ca & 125 & 11735 & 34894 & 26456 & 14.0 & 46405.6 \\
data.gov.uk & 80 & 15272 & 123088 & 72272 & 43.5 & 9428.8 \\
data.gov.au & 32 & 2308 & 21654 & 20107 & 16.5 & 16643.6 \\
data.gov.ie & 11 & 12431 & 36534 & 24668 & 8.3 & 3908.5 \\
africaopendata.org & 6 & 3417 & 42173 & 38493 & 24.4 & 1553.6 \\
data.gov.sg & 2 & 1309 & 2409 & 2336 & 5.3 & 4609.1 \\

\hline
\end{tabular}
\end{table*}

The metadata associated with tables can provide additional ways to organize the data.  For example, the organization that produced the dataset is provided in the CKAN metadata.  Tags, which are keywords describing the topic of the data, are also provided at the level of a dataset.  Figure \ref{fig:ckan_metadata} shows key metadata we extract from the CKAN datasets.

\begin{figure}
    \centering \small
\begin{lstlisting}[language=json,firstnumber=1,numbers=none]
{ 
  "table_id": str,
  |\color{blue}"dataset\_id"|: str,
  |\color{forestgreen}"organization\_id"|: str,
  |\color{orange}"tags"|: List[str],
  "table_name": str,
  "table_description": str,
  "column_headers": [
    {"name": str, "desc": str, "dtype": str}
  ]
}
\end{lstlisting}
    \caption{Metadata associated with CKAN datasets}
    \label{fig:ckan_metadata}
\end{figure}

In addition to the metadata used to construct the table search benchmarks, we also use license metadata to filter the benchmark to data with permissive, open licenses.
All of the CKAN sites we used have a ``default license'', meaning a license that is presumed if the individual data sources do not indicate a license. Table \ref{tab:licenses} shows the default license for each site we collected and the fraction of datasets with the default licenses.  We exclude from our benchmark any license that specifies ``no derivatives'', as well as any license given as ``Other''. We also remove data with non-commercial licenses, both to provide for the widest possible use, and because such data represents a small fraction of the total.



\begin{table}[!htb]
\caption{Default Licenses} 
\label{tab:licenses}
\centering
\begin{tabular}{lll}
\hline
Site & License & \% of Datasets \\
\hline
catalog.data.gov & Public Domain & 93.5\% \\
open.canada.ca & Open Government Licence - Canada & 94.9\% \\
data.gov.uk & UK Open Government Licence (OGL) & 99.6\% \\
data.gov.au & Creative Commons Attribution 3.0 Australia & 83.2\% \\
data.gov.ie & Creative Commons Attribution 4.0 & 97.8\% \\
africaopendata.org & UK Open Government Licence (OGL) & 74.4\% \\
data.gov.sg & Singapore Open Data Licence & 100\% \\
\hline
\end{tabular}
\end{table}

After gathering and filtering the data for open licenses, we extract the structured, tabular data.  The datasets are typically provided as archives of csv or Excel files.  We use pandas to extract the column headers as well as five sample rows.

\usetikzlibrary{shapes,shadows,arrows}

\tikzstyle{decision} = [diamond, draw, fill=white]
\tikzstyle{line} = [draw, -stealth, thick]
\tikzstyle{elli}=[draw, ellipse, fill=gray!20, minimum height=8mm, text width=5em, text centered]
\tikzstyle{block} = [draw, rectangle, fill=white, text width=7em, text centered, minimum height=15mm, node distance=8em]

\begin{figure*}
    \centering
\begin{tikzpicture}
    \node [block,fill=blue!30!white] (download) {Download};
    \node [block,fill=blue!30!white, right of=download, xshift=2em] (extract) {Extract};
    \node [block,fill=blue!30!white, right of=extract, xshift=2em] (dedup) {Deduplicate};
    \node [block,fill=blue!30!white, right of=dedup, xshift=2em] (gt) {Build Ground Truth from Metadata};
    \node [elli, above of=download, yshift=3em] (api) {CKAN API};
    \node [elli, above of=extract, yshift=3em] (zip) {CSV and Excel files};
    \node [elli, above of=dedup, yshift=3em] (json) {Tables with metadata};
    \node [elli, above of=gt, yshift=3em] (deduped) {Tables with metadata};
    \path [line] (api) -- (download);
    \path [line] (download) -- (zip);
    \path [line] (zip) -- (extract);
    \path [line] (extract) -- (json);
    \path [line] (json) -- (dedup);
    \path [line] (dedup) -- (deduped);
    \path [line] (deduped) -- (gt);
\end{tikzpicture}
    \caption{Pipeline for processing CKAN data}
    \label{fig:ckan_pipeline}
\end{figure*}
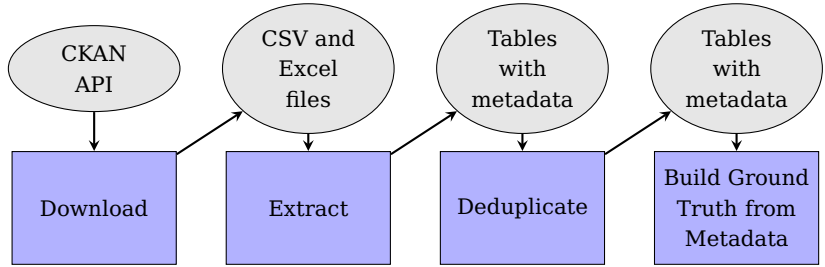

We follow the extraction of the data by de-duplicating the tables, this is needed for the related table search benchmark, since otherwise there will be trivially related, identical tables in the result set.  We took a conservative approach, since part of the benchmark's purpose is verifying that the search system being tested can get the easy cases correct.
If two tables are very similar, such as two tables with the same data but for different years, search methods should be able to identify them as related.
Therefore de-duplication only removes tables as duplicates if the table's names, descriptions, tags, column headers and first five rows are all the same.  Table \ref{tab:duplicates} gives the related table specific pre-processing statistics. 

\begin{table}[!htb]
\caption{Fraction of duplicate tables} 
\label{tab:duplicates}
\centering
\begin{tabular}{rrrrr}
\hline
Site & Distinct & No Data & Duplicate & Fraction Duplicate \\
\hline
catalog.data.gov & 29735 & 27028 & 11322 & 27.58\% \\
open.canada.ca & 25012 & 9610 & 276 & 1.09\% \\
data.gov.uk & 65608 & 53842 & 3679 & 5.31\% \\
data.gov.au & 19449 & 2508 & 530 & 2.65\% \\
data.gov.ie & 24212 & 12082 & 256 & 1.05\% \\
africaopendata.org & 29109 & 4235 & 8881 & 23.38\% \\
data.gov.sg & 2323 & 80 & 6 & 0.26\% \\
\hline
\end{tabular}
\end{table}

The final step is to use the metadata to construct the ground truth for our table search benchmarks.  Figure \ref{fig:ckan_pipeline} illustrates this pipeline.

\section{Table Search}

We consider two types of table search: keyword search and related table search.  These correspond to the initial steps of data discovery, collecting a set of initial tables and expanding this set with related tables.

In table search we consider the tables as documents to be retrieved.  The metadata used to construct the ground truth for the benchmark must not be included in the table representation, but there is still considerable flexibility remaining.  We consider the inclusion of various table elements: table name, table description, column names, and five sample rows of data.  We conduct experiments including different subsets of these table elements to explore their impact on retrieval performance and also to examine the robustness of retrieval systems to data lakes where such information may not be available.

\subsection{Keyword Search}
In keyword search the input is a set of keywords, or key-phrases more generally.  Typical keyword search benchmarks will ask humans to formulate an intent or information need, then write keywords to express this intent.  The results can then be judged as relevant to this intent.  In our benchmark we use the tag metadata to formulate the keyword queries and tables are judged as relevant if they are tagged with the query key-phrases.

Some key-phrases are used to tag a significant fraction of the catalog documents. We filter out all tags used in more than 10\% of the documents. We also filter out key-phrases assigned to less than five documents. The remainder of tags are assumed to be similar to what a user of search may seek.

Next, we consider queries composed from a pair of tagged key-phrases. Search using these pairs should retrieve documents that contain both of these terms. We perform some additional filtering: Some of these key-phrase pairs may be highly correlated in terms of the documents they reference. For example a phrase and its abbreviation may often be used. To avoid adding these correlated terms in the ground truth, we select only key-phrase pairs that point to document sets with less than 0.5 Jaccard score.
Additionally, similar to the single key-phrase queries, we remove key-phrase pairs that refer to more than 1\% of the documents or less than five documents.

Finally, we build queries from three key-phrases just as with key-phrase pairs but by combining single key-phrase queries with two key-phrase queries rather than single key-phrases with itself.

This benchmark is a challenging semantic search over tables.  The table representation indexed for search contains a match for all key-phrases only 16\% and 8\% of the time for two and three key-phrases search queries respectively.  And in 38\% and 21\% of cases the table representation contains \textit{none} of the key-phrases.

\subsection{Related Table Search}
Related table search takes a table as a query and returns semantically related tables.  Many notions of relevance have been used for this task, as described in Section \ref{sec:related_work}.  We again turn to the metadata to describe our ground truth. A result table is considered relevant for a query table by three different standards, which we evaluate as separate metrics.  Dataset relevance judges two tables as related if they were included in the same dataset.  Typically tables are packaged in the same dataset because the organization that published the data considered them parts of the same overall information.  The second notion of relevance is two tables produced by the same organization.  Typically organizations will stay within a field or topic, so this is close to topical relevance.  Finally, the overlap between the semantic tags assigned to tables can serve to measure their relatedness.  To measure this tag similarity we first measure the frequency of each tag, $k$, in the set of all site tables $T$. Then the tag similarity score between two tables $t_i$ and $t_j$ is $\mathcal{S}^{tag}(t_i, t_j)$ given in equation \ref{eq:tag_sim}.

\begin{equation}\label{eq:tag_sim}
\begin{aligned}
freq(k) & = |\{ t \in T : k \in t^{tags} \} |  \\
\mathcal{S}^{tag}(t_i, t_j) & = \sum_{k \in t_i^{tags} \bigcap t_j^{tags}} \log_2 \left( |T| / freq(k) \right)\\
\end{aligned}
\end{equation}


\section{Baselines}

First we apply the conventional (non-neural) information retrieval baseline: BM25~\cite{bm25}, as provided by Anserini~\cite{anserini}.  We test BM25 for related table search over all datasets.  For baselines involving learned retrieval and for keyword search we test only on Canada, using the other sites as training data.  Table \ref{tab:bm25_results} gives the results for related table search over all sites using a table representation of \texttt{table\_name}, \texttt{column\_headers}, and \texttt{sample\_data}.  We report Normalized Discounted Cumulative Gain \cite{ndcg}, considering only the top 20 results (NDCG@20).

\begin{table}[!htb]
\caption{BM25 results for related table search (NDCG@20)} 
\label{tab:bm25_results}
\centering
\begin{tabular}{rccc}
\hline
Site & Dataset & Organization & Tag \\
\hline

catalog.data.gov & 56.1\% & 50.8\% & 48.2\% \\
open.canada.ca & 58.3\% & 60.3\% & 19.8\% \\
data.gov.uk & 54.6\% & 59.0\% & 55.2\% \\
data.gov.au & 46.5\% & 61.3\% & 49.8\% \\
data.gov.ie & 49.8\% & 62.5\% & 52.7\% \\
africaopendata.org & 32.1\% & 63.1\% & 63.3\% \\
data.gov.sg & 55.1\% & 54.6\% & 53.3\% \\

\hline
\end{tabular}
\end{table}

To go beyond the BM25 baseline, we use Neural Information Retrieval.  In this paradigm the tables in the datasets are projected to a dense vector and indexed in an Approximate Nearest Neighbor (ANN) datastructure.  Query tables are also projected to vectors, then the $k$-nearest vectors are located through the ANN. The associated tables are then the results.  

ColBERT~\cite{colbert} is a approach to neural information retrieval based on \textit{late interaction}.  Token vectors for queries and documents are computed independently using a BERT$_{\text{BASE}}$ transformer encoder.  Document token vectors are stored in an ANN and for each query vector the most similar document token vectors are retrieved.  The documents containing the similar token vectors are then the candidates to be scored.  The scoring function takes the most similar document token for each query token and sums the similarity scores over all query tokens.

A typical approach for NIR uses a biencoder architecture, such as Dense Passage Retrieval (DPR)~\cite{dpr}.  DPR was developed for question answering and uses two transformers, each initialized with the BERT$_{\text{BASE}}$ model. One transformer, the query encoder is trained to project the questions to vectors while the other, the context encoder, is trained to project passages from the corpus to vectors.

A biencoder architecture is logical when the queries and candidate results are different sorts of things. However, in related table search the problem is symmetric. Both the queries and results are tables, and the notion of relatedness is symmetric. We therefore use a ``Siamese'' architecture where both the query and context encoder are a single transformer. We initialize this transformer from the context encoder of DPR and further fine-tune it for related table retrieval.








\begin{table}[!htb]
\caption{Related Table Search Results as NDCG@20} 
\label{tab:results}
\centering
\begin{tabular}{l|rrr}
\hline
Method & Dataset &  Organization & Tag \\
\hline
BM25 & 58.3\% & 60.3\% & 19.8\% \\
Siamese & 85.1\% & 94.8\% & 27.8\% \\
BiEncoder & 83.9\% & 94.2\% & 27.1\% \\
\hline
\end{tabular}
\end{table}

Table \ref{tab:results} shows our results for related table search, testing on the Canadian open data and using the other sites as training.  We initially experimenting with training separate models for each of the three relatedness measures, but found a single model performed almost as well.  The neural models, Siamese and BiEncoder, perform better over all three relatedness measures, with the largest difference in the Organization relatedness. The Siamese architecture performs slightly better due to its better fit for a symmetric task.



\begin{table}[!htb]
\caption{Results for keyword search on Canada} 
\label{tab:keyword_results}
\centering
\begin{tabular}{rccc}
\hline
Model & 	Precision@10 &	R-Precision@50 &	NDCG@50 \\
\hline
BM25 & 0.183 &	0.279 &	0.422 \\
ColBERT & 0.207	& 0.321	& 0.472 \\
\hline
\end{tabular}
\end{table}


Table \ref{tab:keyword_results} gives the performance of BM25 and ColBERT on Canadian data.  The ColBERT model is not fine tuned on OGDC, since we found its performance dropped greatly when attempting transfer across sites.  There is a slight gain even from using the pre-trained ColBERT model over BM25 for keyword search.

\section{Conclusion}

We propose Open Government Data Corpus (OGDC), a dataset providing benchmarks for semantic table search.  OGDC allows automatic evaluation of keyword search and related table search systems by using the metadata released with open government data.  Unlike previous table search benchmarks, OGDC uses realistic data tables, rather than tables intended for display.  We evaluated both traditional and neural baselines and find there is still considerable headroom for improving table search.  OGDC is available at 10.5281/zenodo.7908079.
%
%
%
\bibliographystyle{splncs04}
\bibliography{ckan}

\end{document}